

\def\ifundefined#1{\expandafter\ifx\csname
#1\endcsname\relax}

\newcount\eqnumber \eqnumber=0
\def\beq{ \global\advance\eqnumber by 1 $$ }
\def\eeq{ \eqno(\the\eqnumber)$$ }
\def\label#1{\ifundefined{#1}
\expandafter\xdef\csname #1\endcsname{\the\eqnumber}
\else\message{label #1 already in use}\fi}
\def\(#1){(\csname #1\endcsname)}
\def\puteqno{\global\advance \eqnumber by 1 (\the\eqnumber)}

\newcount\refno \refno=0
\def\[#1]{\ifundefined{#1}\advance\refno by 1
\expandafter\xdef\csname #1\endcsname{\the\refno}
\fi[\csname #1\endcsname]}
\def\refis[#1]{\item{\csname #1\endcsname.}}


\baselineskip=18pt
\magnification=1200


\def\cpn{ {{\rm CP}^{N-1}}}
\def\sphere{ {{\rm S}^{2N-1}}}
\def\qed{${\rm QED}_3$}
\def\tr{ {\rm Tr}}
\def\arctg{ {\rm arctg}}
\def\dlmom{ {{d^3 l}\over{(2\pi)^3}}}

\def\dpmom{ {{d^3 p}\over{(2\pi)^3}}}
\def\dkmom{ {{d^3 k}\over{(2\pi)^3}}}


\vskip1in
\centerline{{\bf$\cpn$ MODEL WITH A CHERN-SIMONS TERM}}
\vskip1in
\centerline{G. Ferretti\footnote\dag{Permanent Address: University of
            Rochester, Department of Physics and Astronomy, Rochester
            NY, 14627}}
\centerline{{\sl and}}
\centerline{S.G. Rajeev$\dagger$}
\centerline{{\sl Research Institute for Theoretical Physics}}
\centerline{{\sl Siltavuorenpenger 20 C}}
\centerline{{\sl SF-00170 Helsinki, Finland}}
\vskip1in
\centerline{\bf Abstract}
The $\cpn$ model in three euclidean dimensions is studied in the presence
of a Chern-Simons term using the $1/N$ expansion. The $\beta$-function for
the statistics parameter $\theta$ is found to be zero to order $1/N$
in the unbroken phase by an explicit calculation.
It is argued to be zero to all orders. Some remarks on the $\theta$
dependence of the critical exponents are also made.
\vfill\eject
The $2+1$ dimensional $\cpn$ model (without Chern-Simons term)
has been extensively studied in the last
decade since it was proven to be renormalizable in the $1/N$ expansion
contrary to naive power counting \[arefeva]. Like many other low dimensional
systems, its importance is twofold. First, it provides a theoretical
laboratory for the study of phenomena that are expected to occur in $3+1$
dimensional gauge theories and secondly, it can be interpreted as an
effective theory for the description of condensed matter systems. The
addition of a Chern-Simons (CS) term \[cs] is of
particular interest in this second aspect. Its addition to the
$\cpn$ model has recently been considered by S.H. Park \[park].

Let us first recall the structure of the $\cpn$ model without CS term
\[cpnmodel].
Let $z$ be an $N$ component complex field. The condition for $z$ to
represent a point in $\cpn$ can be implemented by the
introduction of a Lagrange multiplier $\alpha$ ensuring that $|z|={\rm
const.}$ and a dummy abelian gauge field $A$ ensuring that $z\equiv
e^{i\phi}z$.
This leads to the lagrangian:
\beq
  {\cal L}_\cpn=|D_A z|^2+
  \alpha\bigg(|z|^2-{{N\Lambda}\over{g}}\bigg),
  \label{cpnwithoutcs}
\eeq
(The symbol $D_A$ represent the covariant derivative of $z$).
The ``length square'' of $z$ has been chosen to be $N\Lambda/g$ for
later convenience; $\Lambda$ will also play the role of a cut-off and
$g$ is a dimensionless coupling constant.

It is well known that quantum corrections will generate a kinetic term for
the ``photon'' $A$, making it into a truly propagating degree of freedom.
However, field $A$ can be promoted to a propagating field already at the
classical level by simply adding a kinetic term to the lagrangian
\(cpnwithoutcs).
In three dimensions, beside the usual Maxwell term, we have the option of
choosing the CS term:
\beq
  {\cal L}_{{\rm CS}}=i{\theta\over{4\pi}}\epsilon^{\mu\nu\lambda}
  A_\mu\partial_\nu A_\lambda. \label{csterm}
\eeq
Notice that in this model $\theta$ need not be quantized.

The importance of this term is manifest from expression\(csterm). First of
all, it involves only one derivative and it is therefore expected to
dominate the usual Maxwell term at low momenta (large distances). Second,
the presence of the epsilon tensor brakes parity and time reversal
preserving the combination of the two. This is what is expected
in a two dimensional system with an external magnetic field along the
perpendicular direction. Finally, the CS term does not depend on the
metric. Thus, its addition to the effective field theory is insensitive to
the microscopic details that depend on the metric.

We study the system described by ${\cal L}_\cpn+{\cal L}_{{\rm CS}}$ in the
$1/N$ expansion.
As usual \[arefeva], the first step consists in integrating out the
$z$ degrees of freedom, yielding an effective action for $A$ and $\alpha$.
Then, we impose the vanishing of the linear term in the effective action
itself, yielding the ``gap equation'' defining the two phases of the
system:
\beq
    \int^\Lambda\dpmom\;{1\over{p^2+\alpha_c}} - {\Lambda\over{g}}
    +|z_c|^2=0\quad\hbox{and}\quad z_c\alpha_c=0. \label{gapeq}
\eeq
We shall work in the unbroken phase ($g>g_c$), where we assume
$\alpha_c=m^2$, $\sigma=\alpha-\alpha_c$, and $<A_\mu>=0$.
In terms of the shifted field $\sigma$, the ``photon'' field $A$ and the
sources $J$ and $K_\mu$ for $z$ and $A_\mu$, the generating functional
reads:
\beq
    \eqalign{{\cal Z}[J,K]=&\int{\cal D}\sigma{\cal D}A\; \exp\bigg\{
    -N\bigg[\tr\log(-D_A^2+m^2+\sigma)-\{{\rm linear\;\; terms}\}+\int dx\;
    {\cal L}_{{\rm CS}}\bigg]+\cr & \int
    dxdy\;J(x)^*<x|(-D_A^2+m^2+\sigma)^{-1}|y>J(y)+\int dx\;
    K\cdot A\bigg\}.\cr} \label{generating}
\eeq
Notice that we have chosen the coefficient of the CS term to scale like $N$
in order to enhance its contribution in the $1/N$ expansion.

The Feynman rules for the $1/N$ expansion can be obtained by expanding the
trace in powers of the fields \[arefeva], \[cpnmodel].
While doing so, it is convenient to rescale
the fields by a factor $\sqrt{N}$ so that the propagators will all
be independent on $N$ and the vertices will all scale like
inverse powers of $N$. In particular, the $\sigma$-propagator,
represented by a dashed line throughout the paper, is given by:
\beq
    \Sigma(p)=-{{4\pi |p|}\over{\arctg(|p|/2m)}}, \label{sigmaprop}
\eeq
where the peculiar minus sign is a reminder of the fact that $\sigma$
is actually a Lagrange multiplier.
The photon propagator, represented by a curly line is given by:
\beq
    D_{\mu\nu}(p)=\bigg[{{\Gamma(p)}\over{\Gamma^2(p)+\left({{\theta}
    \over{4\pi}}\right)^2 p^2}}\big(\delta_{\mu\nu}-{{p_\mu p_\nu}
    \over{p^2}}\big)+{{\theta/4\pi}\over{\Gamma^2(p)+\left({{\theta}
    \over{4\pi}}\right)^2 p^2}}\epsilon_{\mu\nu\rho}p^\rho\bigg],
    \label{photon}
\eeq
with
\beq
    \Gamma(p)={{p^2+4m^2}\over{8\pi |p|}}\arctg(|p|/2m)-{m\over{4\pi}}.
\eeq
When needed, the propagator for the $z$ field itself,
\beq
    \Delta(p)={1\over{p^2+m^2}}
\eeq
will be represented by a solid line.

The first question we would like to address is whether or not the statistics
parameter $\theta$ is renormalized in the $1/N$ expansion. At first glance
it seems like the Coleman-Hill theorem \[coleman]
is not directly applicable here since its proof relies on perturbative
arguments. There are known cases where extra corrections can
appear\[hagen]. However, we will show that this is not the case in this
model, i.e., the statistics parameter is not renormalized.

Let us consider the $1/N$ correction to the photon
(inverse) propagator. In terms of
the non-local vertices defined by the expansion of \(generating), this can
be represented by the five (amputated)
diagrams in fig. $(1)$. Of course, each diagram
represents the sum of many local diagrams constructed by substituting
$z$-loops to each vertex and attaching to it the propagators in all
possible ways.

We write the contribution of each diagram to the inverse propagator as
\beq
    \Pi_{\mu\nu}^{(i)}(p)=\big(\delta_{\mu\nu}-{{p_\mu p_\nu}
    \over{p^2}}\big)\Pi_S^{(i)}(p)-{\theta\over{4\pi}}\Pi_A^{(i)}(p)
    \epsilon_{\mu\nu\rho}p^\rho\qquad i=1\cdots 5. \label{corrections}
\eeq
We are interested is the zero momentum limit of the anti-symmetric part
$\Pi_A=\sum \Pi_A^{(i)}$, the renormalization constant for $\theta$
being given by $Z_\theta=1+\Pi_A(0)$.

Diagrams $1$, $2$ and $3$ yield manifestly symmetric expressions, i.e.,
for $i=1,2$ and $3$, $\Pi_A^{(i)}(p)$ is
identically zero for all values of $p$ and
we need not consider them any further.

Diagram $4$ gives
\beq
    \Pi_{\mu\nu}^{(4)}(q)={2\over N}\int\dlmom v_{\mu\sigma}(q,l)
    v_{\nu\tau}(q,l)D_{\sigma\tau}(l)\Sigma(l-q), \label{pifour}
\eeq
where $v$ represents the non-local vertex:
\beq
     v_{\mu\sigma}(q,l)=\int\dpmom\bigg[-\delta_{\mu\sigma}\Delta(p+l)
     \Delta(p+q)+(2p_\mu+q_\mu)(2p_\sigma+l_\sigma)\Delta(p)
     \Delta(p+l)\Delta(p+q)\bigg].
\eeq

It is a simple matter to check that $v$ satisfies the relations
$q_\mu v_{\mu\sigma}(q,l)=v_{\mu\sigma}(q,l) l_\sigma =0$ and
$v_{\mu\sigma}(q,l)=v_{\sigma\mu}(l,q)$ and that it can be written as:
\beq
    v_{\mu\sigma}(q,l)=\big(\delta_{\mu\sigma}(l\cdot q) -
    l_\mu q_\sigma\big)\phi(q,l) + \big(l^2q^2\delta_{\mu\sigma}
    -q_\mu q_\sigma l^2 - l_\mu l_\sigma q^2+q_\mu l_\sigma l\cdot q\big)
    \psi(q,l), \label{vertexv}
\eeq
where the two scalar functions in \(vertexv) have Feynman parametrization:
\beq
    \phi(q,l)=-{1\over{4\pi}}\int_0^1d\alpha\;\int_0^{1-\alpha}d\beta\;
    {{\alpha\beta}\over{(m^2+\alpha q^2+\beta l^2 -
    (\alpha q+\beta l)^2)^{3/2}}}, \label{phiscalar}
\eeq
and
\beq
    l^2\psi(q,l)={1\over{8\pi}}\int_0^1d\alpha\;\int_0^{1-\alpha}d\beta\;
    {{\alpha-2\alpha^2}\over{(m^2+\alpha q^2+\beta l^2 -
    (\alpha q+\beta l)^2)^{3/2}}}. \label{psiscalar}
\eeq

The anti-symmetric part of $\Pi^{(4)}_{\mu\nu}$ can be extracted in the usual
way, by contracting with the epsilon tensor. In our notations:
\beq
    \Pi_A^{(4)}(q)=-{1\over N}\int\dlmom\;{{\Sigma(l-q)}\over{\Gamma^2(l)
    +\left({{\theta}\over{4\pi}}\right)^2 l^2}}\bigg[2l\cdot q l^2
    \phi^2+2l^2(l^2q^2+(l\cdot q)^2)\phi\psi+2(l\cdot q)q^2l^4\psi^2
    \bigg]. \label{piafour}
\eeq
The functions $\phi$ and $\psi$ are analytic at $q=0$ and their asymptotic
behaviour for large $l$ is such that the integral in\(piafour) converges.
Hence, $\Pi_A^{(4)}(0)=0$ and there is no renormalization for $\theta$
coming from the fourth diagram in fig. $(1)$.

Diagram number $(5)$ can be treated similarly. It can be written in compact
form as:
\beq
    \Pi_{\mu\nu}^{(5)}={1\over N}\int\dkmom\;D_{\rho\lambda}(k)
    v_{\mu\nu\rho\lambda}(k,q), \label{pifive}
\eeq
where $v_{\mu\nu\rho\lambda}(k,q)$ is the highly non-local expression
representing the four photon vertex. Since there are no $\sigma$-lines in
this diagram, this is the same expression that we would obtain by summing
all the two loop diagrams of scalar \qed \[semenoff]. In other words,
the only substantial difference between
the expression of $\Pi_{\mu\nu}^{(5)}$ and the
analogous quantity of scalar \qed is in the form of the internal photon
propagator $D_{\rho\lambda}(k)$. But the expression for the vertex alone is
enough to guarantee a vanishing contribution to the renormalization of
$\theta$, i.e., without having to perform the integral over $k$ in
\(pifive) it is easy to check that
\beq
    \Pi_A^{(5)}(0)=\bigg[-{{2\pi}\over{3\theta}}\epsilon_{\mu\nu\tau}
    {\partial\over{\partial q_\tau}}\Pi_{\mu\nu}^{(5)}(q)\bigg]_{q\to 0}=0.
    \label{piafive}
\eeq
This was the last possible source of divergence to order $1/N$, so we have
explicitly checked that there is no renormalization of $\theta$ to this
order.

We now argue that there is no renormalization for $\theta$ to all orders
in $1/N$. This is an extension of the well-known Coleman-Hill theorem
\[coleman] to the $1/N$ expansion. Let us briefly recall how the theorem is
proven in ordinary perturbation theory. We restrict ourselves to the case
of scalar \qed with CS term \[semenoff].

The most general diagram that appears in the correction for the photon
propagator $\Pi_{\mu\nu}(q)$ is made of two external photon lines and an
arbitrary number of bosonic loops and internal photon lines. Each bosonic
loop defines a non-local vertex for the photons. Consider a particular
vertex $\Gamma_{\mu_1\mu_2\cdots\mu_m}(k_1,k_2,\cdots,k_m)$ with $m$
incoming photon lines of momentum $k_1,\cdots,k_{m-1}$ and
$k_m=\sum_1^{m-1} k_i$. By gauge invariance and analyticity of $\Gamma$, it
is easy to see that
\beq
    \Gamma_{\mu_1\mu_2\cdots\mu_m}(k_1,k_2,\cdots,k_m)=
    O(|k_1||k_2|\cdots|k_{m-1}|). \label{ogamma}
\eeq
We have to consider two possible cases. In the first case, the two external
photon legs (carrying momentum $q$) are attached to two different bosonic
loops. In this case, by applying \(ogamma) twice, we obtain
$\Pi_{\mu\nu}(q)=O(q^2)$. From this fact and \(piafive)
$\Pi_A(0)=0$ immediately follows.

In the second case, both external legs are attached to the same loop and
\(ogamma) also implies $\Pi_{\mu\nu}(q)=O(q^2)$ {\it unless} $m=2$.
But the case $m=2$ corresponds to the one loop diagrams, for which it can
be explicitly checked that the anti-symmetric part
$\Pi_A$ vanishes identically. This ends our sketchy review of the
Coleman-Hill theorem.

In our model the vertices are even better behaved. For example, vertices
with only two incoming photon lines are forbidden in the sense of the large
$N$ limit \[arefeva]. The only two types of vertex involving photons are
either those with two photon lines and $r\geq 1$ $\sigma$-lines:
\beq
    \Gamma_{\mu_1\mu_2}(k_1,k_2,p_1,\cdots p_r)\qquad
    p_r=k_1+k_2+\sum_1^{r-1} p_i, \label{firsttype}
\eeq
or those with $m\geq 3$ photons and $s\geq 0$ $\sigma$-lines
\beq
    \Gamma_{\mu_1\mu_2\cdots\mu_m}(k_1,k_2,\cdots,k_m,p_1,\cdots p_r),
    \label{secondtype}
\eeq
(the last momentum in \(secondtype) is also fixed by momentum conservation).
But for both types of vertex \(firsttype) and \(secondtype) it is now true
that
\beq
    \Gamma=O(|k_1||k_2|)
\eeq
and therefore, any diagram involving only two external photon legs of
momentum $q$ is $O(q^2)$, yielding again $\Pi_A(0)=0$.

This proves the validity of the Coleman-Hill theorem in the $1/N$
expansion. If there is any non-perturbative effect that invalidates the
theorem, it must be non perturbative in $1/N$ as well as in the coupling
constant. The result can also be generalized to the fermionic case.
Recall that in the presence of fermions there are non zero radiative
corrections coming from one-loop graphs in ordinary perturbation theory.
In the large $N$ limit, the same correction will be present in
the photon propagator at leading order in $1/N$. There will be no
corrections coming from lower orders as long as we stay in the analyticity
region.

Now that we have established the non-renormalization of the statistics
parameter $\theta$ in the $1/N$ expansion it makes sense to calculate the
dependence of the critical exponents on $\theta$ itself.
This was first done in \[park]. There, dimensional regularization was used
throughout the calculation and the critical exponent $\nu$ was obtained by
studying the $zz\sigma$ three-point function. This may cause some concern
because of the presence of an epsilon tensor that is a truly three
dimensional object and because of the fact that $\sigma$ is not a
propagating field, so it is a little distressing to consider external
$\sigma$-lines. We recover the same results by following a slightly
different approach that does not use either assumptions.

We use gauge invariant projection with an explicit cut-off $\Lambda$. The
critical exponents $\eta$ and $\nu$ are both calculated from the inverse
$z$ propagator. The relevant diagrams to order $1/N$ are shown in
fig. $(2)$. We have explicitly included all the tadpole diagrams that
arise to this order. Also, we have chosen to draw the contribution in
terms of local diagrams, (i.e., with the $z$-loops explicitly indicated)
because their number is not as large as it was for the photon propagator.
One can always recover the previous picture by shrinking each
$z$-loop into a non-local vertex. In this case, the last two diagrams of
fig. $(2)$ would be combined into a single one.

Applying the ordinary Feynman rules and evaluating the diagrams
asymptotically, we obtain the following expression for the divergent
part of the $z$ propagator: ($\bar\theta=4\theta/\pi$)
\beq
    \tilde\Delta^{-1}(p)={1\over N}\bigg[-{{20}\over{\pi^2}}\bigg(
    1-{{16}\over{15}}{{\bar\theta^2}\over{1+\bar\theta^2}}\bigg)\log
    (\Lambda/m)p^2+{{76}\over{\pi^2}}\bigg(1-{{64}\over{76}}
    {{\bar\theta^2(\bar\theta^2+5)}\over{(1+\bar\theta^2)^2}}\bigg)
    \log(\Lambda/m)m^2\bigg]. \label{zzfirstorder}
\eeq
Actually, a naive evaluation of
the diagrams in fig. $(2)$ will give rise to a
linear divergence in the mass term. But this divergence simply redefines
the (non universal) critical coupling constant $g_c$ and does not affect
any of the universal quantities.

{}From \(zzfirstorder) ore can immediately read off the critical exponents:
\beq
    \eqalign{\eta=&-{{20}\over{\pi^2N}}\bigg(1-{{16}\over{15}}
    {{\bar\theta^2}\over{1+\bar\theta^2}}\bigg)\cr
    \nu=&1-{{48}\over{\pi^2N}}\bigg(1-{8\over 9}{{\bar\theta^2
    (\bar\theta^2+4)}\over{(1+\bar\theta^2)^2}}\bigg) .\cr} \label{crit}
\eeq
Setting $\theta=0$ we recover the usual $\cpn$ result
$\eta_\cpn=-20/\pi^2N$ and $\nu_\cpn=1-48/\pi^2N$, whereas, taking the limit
$\theta\to\infty$, we obtain the critical exponents for the $\sphere$ model
$\eta_\sphere=4/3\pi^2N$ and $\nu_\sphere=1-16/3\pi^2N$.
This last result
is simply explained by recalling that the CS term acts as a mass term for
the photon and by letting $\theta$ go to infinity we simply remove the
photon from the spectrum, leaving only the constraint $|z|={\rm const.}$,
which is just the $\sphere$ model.

This fact, together with the non renormalization of $\theta$ raises
some interesting issues. We have showed that to all orders in $1/N$ there is
a line of fixed points, described by $g=g_c$ and $\theta$, that
interpolates between the ordinary $\cpn$ model without CS term and the
$\sphere$ model. It would be interesting to investigate whether this line
is ``broken'' by some phenomenon that is non-perturbative in $1/N$. This
can be answered by studying the object equivalent to the equation of
motion, which, in our case is the gap equation describing the two phases.

Another point worth studying is what happens at the critical point, where
the photon propagator is no longer analytic. According to the general
argument presented in \[semenoff], the $\beta$-function for $\theta$ will
still be zero but we expect a finite renormalization to occur. This
point, however, has to be studied by numerical analysis since already in
the simpler setting of \qed it has proven impossible to
evaluate the Feynman integrals.

We would like to thank the Research Institute of Theoretical Physics of
Helsinki for the hospitality. This work was supported in part by the
DOE contract No. DE-AC-02-76ER13065.

\vfill\eject

\centerline{{\bf References}}

\refis[arefeva] I.Ya. Aref'eva and S.I. Azakov, Nucl. Phys. B162 (1980) 298.
\refis[cs] S. Deser, R. Jackiw and S. Templeton, Ann. Phys. 140 (1982) 372.
\refis[park] S.H. Park, ICTP Preprint IC/91/5 (1991).
\refis[cpnmodel] H. Eichenherr, Nucl. Phys. B146 (1978) 215.
                 \hfill\break
                 A. D'Adda, M. L\"uscher and P. Di Vecchia, Nucl. Phys.
                 B146 (1978) 63.
                 \hfill\break
                 V.L. Golo and A.M. Perelomov, Phys. Lett. B79 (1978) 112.
\refis[coleman]  S. Coleman and B. Hill, Phys. Lett. B159 (1985) 184.
\refis[hagen] C. Hagen, P. Panigrahi and S. Ramaswamy, Phys. Rev. Lett.
              61 (1988) 389.
\refis[semenoff] G.W. Semenoff, P. Sodano and Y.S. Wu, Phys. Rev. Lett.
              62 (1989) 715.

\vfill\eject

\centerline{{\bf Figure Captions}}

\centerline{Fig. $(1)$: Corrections to the photon propagator to order $1/N$.}

\centerline{Fig. $(2)$: Corrections to the $z$ propagator to order $1/N$.}

\bye